\documentclass[preprint]{revtex4}
\usepackage{amssymb,amsmath}
\usepackage{graphicx,float}
\usepackage{indentfirst}

\begin{document}

\title{Static Domain Wall in the Braneworld gravity}
\author{M. C. B. Abdalla}
\email{mabdalla@ift.unesp.br}
\affiliation{Instituto de F\'isica Te\'orica, UNESP - Universidade Estadual Paulista,
Rua Dr. Bento Teobaldo Ferraz 271, Bloco II, Barra-Funda,
Caixa Postal 70532-2, 01156-970, S\~ao Paulo, SP, Brazil.}
\author{P. F. Carlesso}
\email{pablofisico@ift.unesp.br}
\affiliation{Instituto de F\'isica Te\'orica, UNESP - Universidade Estadual Paulista,
Rua Dr. Bento Teobaldo Ferraz 271, Bloco II, Barra-Funda,
Caixa Postal 70532-2, 01156-970, S\~ao Paulo, SP, Brazil.}
\author{J. M. Hoff da Silva}
\email{hoff@feg.unesp.br; hoff@ift.unesp.br}
\affiliation{Departamento de F\1sica e Qu\1mica, UNESP - Universidade
Estadual Paulista, Av. Dr. Ariberto Pereira da Cunha, 333,
Guaratinguet\'a, SP, Brazil.}
\begin{abstract}
In this paper we consider a static domain wall inside a 3-brane. Differently of the standard achievement obtained in General Relativity, the analysis performed here gives a consistency condition for the existence of static domain walls in a braneworld gravitational scenario. It is also shown the behavior of the domain wall gravitational field in the newtonian limit.
\end{abstract}
\maketitle
\section{Introduction}
\noindent Braneworld models have been substantially studied in contemporaneous theoretical physics. Some of the reasons for that are the seminal works by Randall-Sundrum \cite{rs1,rs2}. For instance, in Ref. \cite{rs1} it is provided a possible explanation for the hierarchy problem of particle physics. These models are partially inspired on results from strings theory, that necessarily require the existence of extra-dimensions. The braneworld picture may be understood as an effective scenario of Horava-Witten framework \cite{witten}, within a warped spacetime. It is interesting to analyse the physical implications of extra dimensions, whose consequences can include particle physics effects, new astrophysics observables \cite{MAAR}, and some cosmological modifications from the standard model. For example, the dark matter problem, raised in the realm of astrophysics and cosmology set up, requires particles with no electromagnetic and strong interactions. This last problem can be approached via braneworld scenarios, in which dark matter can be interpreted as massive gravitons from extra dimensions \cite{RM,mak}. Nevertheless, there are many other situations where braneworlds can evoke new insights about physical predictions. Some of those situations are the topological defects in cosmology. Those defects may be generated by means of one or several spontaneous symmetry breaking in the lagrangian of some scalar field models. In this process, structures as domain walls and cosmic strings may be created at cosmological scale \cite{zeldovich,kibble}. There are several works studying the gravitational properties of domain walls. Some exhaustive achievements were founded for the domain wall behavior in the context of usual general relativity theory \cite{vilenkin,ipser,widrow}.

A relevant result, which we are particularly interested in, shows that thick and static domain walls are incompatible with general relativity \cite{ray}. It is an appealing and strong result constituting a benchmark in the study of gravitational effects of domain walls. In this work we shall revisit this result in the light of braneworld gravity. In fact, the study of domain walls gravitational effects within the scope of braneworlds was previously considered  \cite{gregory,silveira,larsen,padilla1,padilla2}. Particularly in the works \cite{padilla1,padilla2}, the authors assume a conformal bulk metric, bringing no significant differences on the domain wall gravitational field from the usual four dimensional Einstein equation.

Assuming that general relativity holds in the whole five dimensional bulk, and the four dimensional brane is endowed with $\mathbb{Z}_2$ symmetry, one can recover the gravitational equation on the brane in a suitable and precise way. The result stands for corrections coming from the extra dimension (encoded in the Weyl tensor), as well as some modifications proportional to the square of the stress tensor \cite{jap,ali}. For the specific case investigated in this paper, the contributions from the square of the stress tensor are identically zero, and all the modifications rest upon the Weyl tensor term. As we shall see, the Weyl term, under very general assumptions, is the responsible for allowing the existence of static domain walls in the braneworld. This result is in acute contrast with the one obtained within usual general relativity.

This paper is organized as follows: in Section II, we briefly review the main steps leading to an effective (and modified) gravitational field equation on the brane. Going further, we show that static domain walls are allowed in the braneworld gravity. In the final Section we conclude, giving a simple physical interpretation and pointing out some perspectives in this branch of research.

\section{braneworld gravity}

As previously mentioned, the extra-dimension model considered here is quite based on the Randall-Sundrum models (mostly  the paper \cite{rs2}). Therefore, we consider a 3-brane, in which the standard model particles are localized, embedded in a 5-dimensional bulk with a $\mathbb{Z}_2$ symmetric fifth dimension. That scenario, provides a distinct description for the gravitational dynamics. To achieve the gravitational equations on the brane, it is taken for granted that Einstein's field equations hold in 5 dimensions. Then, throughout Guass-Codazzi formalism and Israel matching conditions it is possible to obtain the gravitational equations \cite{jap,bonjour} relating the intrinsic brane quantities (metric $g_{\mu\nu}$ and Einstein's $G_{\mu\nu}$ tensors) with the brane stress-energy tensor $\tau_{\mu\nu}$
\begin{equation}\label{dyn}
G_{\mu\nu}=-\Lambda g_{\mu\nu}+\kappa_{4}^{2}\tau_{\mu\nu}+\kappa_{5}^{4}\Pi_{\mu\nu}-E_{\mu\nu},
\end{equation}
where $\Lambda=\frac{\kappa_{5}^{2}}{2}(\Lambda_{5}+\kappa_{5}^{2}\lambda_{b}^{2}/6)$ is an effective brane cosmological constant (determined by the bulk cosmological constant $\Lambda_{5}$ and the intrinsic brane tension $\lambda_{b}$). These equations are closely related with the Einstein's field equations, the distinction is performed by the two last terms on the right-hand side.  The tensor $\Pi_{\mu\nu}$ is given by
\begin{equation}\label{s}
  \Pi_{\mu\nu}= -\frac{1}{4}\tau_\mu^\sigma\tau_{\nu\sigma}+\frac{1}{12}\tau\tau_{\mu\nu}+ \frac{1}{8}q_{\mu\nu}\tau_{\alpha\beta}\tau^{\alpha\beta}-\frac{1}{24}q_{\mu\nu}\tau^2,
\end{equation}
which is fully dependent of the brane stress-energy tensor. Otherwise, the $E_{\mu\nu}$ (the so-called Weyl fluid) is a non-local term, depending on the bulk's Weyl tensor. As we shall see, it is this tensor the responsible to enable the existence of static domain walls in the braneworld gravity.

\section{Static Domain Walls}
Let us start outlining the basic formalism concerning domain walls. These structures can be generated by a scalar field lagrangian like as
\begin{equation} \label{lagrangeana}
L=\frac{1}{2}\partial_{\mu}\phi\partial^{\mu}\phi-\lambda^2(\phi^2-\eta^2)^2.
\end{equation}
Its originate a topological anomaly on the transition layer between the two vacuum states associated with the potential of the above lagrangian. If we consider a dependence $\phi=\phi(x)$ (where $x$ is parameterizing one of the spatial dimensions) the classical field equations give us
\begin{equation} \label{eq4}
\phi(x)=\eta\tanh(\sqrt{2}\lambda\eta x).
\end{equation}

By assuming the spacetime nearly Minkowskian, we can determine the stress-energy tensor for the scalar field lagrangian (\ref{lagrangeana})
 \begin{equation}\label{t}
   T_{\mu\nu}=\partial_{\mu}\phi\partial_{\nu}\phi-
\eta_{\mu\nu}\left[\frac{1}{2}\partial_{\alpha}\phi\partial^{\alpha}\phi-\lambda^2(\phi^2-\eta^2)^2\right].
 \end{equation}
 For the scalar field given by (\ref{eq4}), we compute the stress-energy components, so that we rewrite the stress-energy tensor as
\begin{equation} \label{eq13}
T^{\mu}_{\nu}=\sigma(x)(\delta^{\mu}_{\nu}+\xi^\mu\xi_\nu),
\end{equation}
where $\xi^\mu$ is a unity spacelike vector orthogonal to the wall surface and
\begin{equation} \label{eq10}
\sigma(x)=2\lambda^2\eta^4\left[\cosh\left(\sqrt{2}\lambda\eta x)\right)\right]^{-4}.
\end{equation}
The above relations show how the energy for the scalar field is distributed in the spacetime. The function $\sigma(x)$ has a peak centered in $x=0$, characterizing the domains wall, whose thickness is determined by the relation $\delta\sim\frac{1}{\lambda\eta}$.

Many works dealing with domains walls in general relativity consider the thin case limit, utilizing the Dirac delta function to localize the domain wall \cite{vilenkin, ipser}. In this work, we consider a thick domain wall, where the energy distribution in spacetime is performed by (\ref{eq13}) along with (\ref{eq10}).

As seen in the expression (\ref{eq13}), the contraction of $\xi^\mu$ with $T^{\mu}_{\nu}$ is null. Therefore, contracting equation (\ref{t}) with $\xi^\mu$, gives
\begin{equation} \label{eq15}
\xi^\mu\partial_\mu\phi\partial_\nu\phi-\xi_\nu\left[\frac{1}{2}\partial_{\mu}\phi\partial^{\mu}\phi- \lambda^2(\phi^2-\eta^2)^2\right]=0.
\end{equation}
The above expression, shows  that $\partial_\nu\phi$ is proportional to $\xi_\nu$ and therefore its a hypersurface orthogonal vector implying the relation $\nabla_\mu\xi_\nu-\nabla_\nu\xi_\mu=0$. By comparing the expressions for $T^{\mu}_{\nu}$, (\ref{t}) and (\ref{eq13}), we obtain the relation
\begin{equation} \label{eq14}
\sigma=2\lambda^2(\phi^2-\eta^2)^2.
\end{equation}
Taking the partial derivative in the above equation, we conclude that
\begin{equation} \label{eq16}
\partial_\alpha\sigma=\mathcal{N}\xi_\alpha,
\end{equation}
where $\mathcal{N}$ is a scalar.
The energy conservation condition over $T^{\mu}_{\nu}$ in equation (\ref{eq13}) gives
\begin{equation} \label{eq18}
\nabla_\nu\sigma+\nabla_\mu\xi^\mu\xi_\nu+\sigma\nabla_\mu\xi^\mu+\sigma\xi^\mu\nabla_\mu\xi_\nu=0.
\end{equation}
 By contracting the above equation with $\xi^\nu$ and take in account that her is a unit vector (which implies $\xi^\nu\nabla_\mu\xi_\nu=0$), we have
\begin{equation} \label{eq18}
\nabla_\mu\xi^\mu=0.
\end{equation}
Written the Riemann tensor as,
\begin{equation} \label{eq21}
\nabla_\alpha\nabla_\beta\xi^\mu-\nabla_\beta\nabla_\alpha\xi^\mu={R^\mu}_{\lambda\alpha\beta}\xi^\lambda,
\end{equation}
we can use the expression (\ref{eq18}) into the above equation to obtain an relation for the Ricci tensor
\begin{equation} \label{eq24}
S_{\alpha\beta}S^{\alpha\beta}+R_{\alpha\beta}\xi^\alpha\xi^\beta=0,
\end{equation}
where
\begin{equation} \label{eq25}
S_{\alpha\beta}\equiv\nabla_\alpha\xi_\beta.
\end{equation}

The braneworld gravitation, given by means of the expression (\ref{dyn}), allow us to determine an expression for the Ricci tensor. Meanwhile, at first we compute  $\Pi_{\mu\nu}$ (\ref{s}) term, by means of the stress-energy tensor for the domain wall spacetime. In can be readily verified that the result is $\Pi_{\mu\nu}=0$ and therefore, we only have
\begin{equation}\label{dyn2}
 R_{\mu\nu}-\frac{1}{2}g_{\mu\nu}R=8\pi T_{\mu\nu}-\Lambda g_{\mu\nu} -E_{\mu\nu}.
\end{equation}
Taking this relation on the expression (\ref{eq24}), we find the consistency condition for the static domain walls in the braneworld gravitation
\begin{equation} \label{eq56}
S_{\alpha\beta}S^{\alpha\beta}+12\pi\sigma=\Lambda + E_{\mu\nu}\xi^\mu\xi^\nu.
\end{equation}
Now, we perform an analysis about this consistency condition, in two parts.

\subsection{The General Relativity case}
At first step, we consider the absence of the cosmological constant term as well as the contribution from the extra dimension $E_{\mu\nu}\xi^\mu\xi^\nu$. In this case, the equation  (\ref{eq56}) becomes
\begin{equation} \label{eq57}
S_{\alpha\beta}S^{\alpha\beta}+12\pi\sigma=0,
\end{equation}
that is, we recover the consistency condition form the standard general relativity without cosmological constant \cite{ray}. Once $\sigma>0$, it is necessary that $S_{\alpha\beta}S^{\alpha\beta}<0$ to satisfy the condition for the existence of static domain walls is general relativity.

However, to the product $S_{\alpha\beta}S^{\alpha\beta}$ holds  negative valued, it is necessary that $S_{\alpha\beta}$ have complex eigenvalues \cite{eisenhart}. This fact implies a pair of complex conjugate eigenvectors. However, we know that $S_{\alpha\beta}$ has at least two real eigenvalues (because $S_{\alpha}^{\alpha}=0$ and $S_{\alpha\beta}\xi^\alpha=0$).

The condition of staticity for the domain wall implies the existence of a timelike hypersurface orthogonal Killing vector. The Lie derivative over the stress-energy tensor  (\ref{eq13}) along the Killing vector field vanishes, implying
\begin{equation} \label{eq37}
\mathcal{L}_K(3\sigma)=0 \rightarrow K^\alpha\partial_\alpha\sigma=0,
\end{equation}
and
\begin{equation} \label{eq39}
K^\mu\nabla_\mu\xi_\beta=\xi^\mu\nabla_\mu K_\beta.
\end{equation}
Taking into account Eq.  (\ref{eq16}) in (\ref{eq37}), we have
\begin{equation} \label{eq40}
\xi^\alpha K_\alpha=0.
\end{equation}
Hypersurface orthogonality implies
\begin{equation} \label{eq41}
K_{[\alpha}\nabla_\gamma K_{\beta]}=0,
\end{equation}
or yet
\begin{equation} \label{eq42}
\xi^\alpha\ K_{[\alpha}\nabla_\gamma K_{\beta]}=0.
\end{equation}
By means of equations  (\ref{eq39}) and (\ref{eq40}), we obtain
\begin{equation} \label{eq44}
K_{\beta}K^\alpha\nabla_\alpha\xi_\gamma-K_\gamma K^\alpha\nabla_\alpha\xi_\beta=0,
\end{equation}
or, contracting with  $K^\beta$,
\begin{equation} \label{eq45}
K^\alpha S_{\alpha\gamma}=\rho K_\gamma,
\end{equation}
where $\rho=(K^\alpha K^\beta\nabla_\alpha K_\beta)/K^\beta K_\beta$.

The equation  (\ref{eq45}) shows that $K^\alpha$ is an eigenvector of $S_{\alpha\gamma}$. This fact precludes the existence of a pair of complex conjugate eigenvectors for $S_{\alpha\gamma}$, thus the consistency condition (\ref{eq57})  is not satisfied and models of static domain walls are incompatible with  general relativity \cite{ray}.

It is noteworthy that author's in \cite{ray} have not considered the influence of cosmological constant on Einstein's equations to obtain this result (perhaps because the cosmological constant was just a theoretical hypothesis when that article was written). If now we consider the cosmological constant influence, a new condition arises
\begin{equation} \label{lamb}
S_{\alpha\beta}S^{\alpha\beta}+12\pi\sigma=\Lambda.
\end{equation}
As both terms on the left-hand side of the above equation are positive, it is necessary that both have the same order of the cosmological constant, or $\lambda^2\eta^4\sim\Lambda$, leading to a very large thickness for the domain wall.

\subsection{The Braneworld case}
The relation (\ref{eq56}) obtained above constrains the Weyl fluid term. In order to clarify the influence of the Weyl fluid over the domain wall, we write it here in a cosmological fluid form
\begin{equation}\label{eq:c6}
E_{\mu\nu}=-k^4[U (u_\mu u_\nu-\frac{1}{3}h_{\mu\nu})+P_{\mu\nu} +Q_{(\mu} u_{\nu)} ],
\end{equation}
where we decompose the metric tensor by means of a 4-velocity field ($g_{\mu\nu}=h_{\mu\nu}+u_\mu u_\nu$). $U=-k^{-4}E_{\mu\nu}u^\mu u^\nu$ represents the dark radiation component, $P_{\mu\nu}=-k^{-4}[h_{(\mu}^\alpha h_{\nu)}^\beta-\frac{1}{3}h_{\mu\nu}h^{\alpha\beta}]E_{\alpha\beta}$ is the anisotropic pressure and $Q_{\mu}=-k^{-4}h_\mu^\alpha E_{\alpha\beta}u^\beta$ the energy flux associate to the Weyl dark fluid.

The staticity condition for the domain wall spacetime requires a null energy flux. Considering the case of a planar domain wall, to simplify the analysis, $S_{\mu\nu}$ vanishes and the constraint (\ref{eq56}) becomes
\begin{equation} \label{eq61}
12\pi\sigma=E_{\mu\nu}\xi^\mu\xi^\nu.
\end{equation}
Unlike of consistency condition given from general relativity (\ref{eq24}), the equations (\ref{eq56}) and (\ref{eq61}) from the braneworld scenario, are not necessarily  in conflict with the presence of static domain walls. Those relations restrain just the Weyl fluid components along the transversal direction to the domain wall. In terms of dark fluid components, equation (\ref{eq61}) gives
\begin{equation} \label{eq58}
P_{\mu\nu}\xi^\mu\xi^\nu=\frac{1}{3}U(x)-12\pi k^{-4}\sigma(x).
\end{equation}
Plugging this relation into (\ref{eq:c6}), we obtain
\begin{equation}\label{eq:c6}
E_{\mu\nu}=-k^4[U (u_\mu u_\nu-\frac{1}{3}w_{\mu\nu})+12\pi k^{-4}\sigma(x)\xi_\mu\xi_\nu+w_\mu^\alpha P_{\alpha\nu}],
\end{equation}
where $w_{\mu\nu}$ is orthogonal to  $u^\mu$ and $\xi^\mu$ related by ($g_{\mu\nu}=w_{\mu\nu}+u_\mu u_\nu-\xi_\mu \xi_\nu$).

Let us analyze the immediate consequences of this model. For a stress-energy tensor of the form $T_{\mu}^{\nu}=diag(\rho-P_1-P_2-P_3)$, the newtonian limit over Einstein's equations give
\begin{equation}\label{poisson}
\nabla^2\phi=4\pi(\rho +P_1+P_2+P_3).
\end{equation}
An usual attractive matter distribution in spacetime, in the newtonian limit, is described by the Poisson's equation $\nabla^2\phi=4\pi\rho$, where the pressure terms are neglected. In that model, taking the diagonal contribution of dark fluid components together with domain wall stress-energy tensor, we obtain
\begin{equation}\label{poisson2}
\nabla^2\phi=4\pi(2k^4U-\sigma).
\end{equation}
In the absence of dark fluid components, we recover the result of Einstein's gravitation in the newtonian limit \cite{vilenkin}, in which the domain wall exhibits a repulsive gravitational force over test particles. Nevertheless, in the braneworld scenario, the Weyl fluid brings other possibilities for the domain wall gravitation. Depending of the value for the dark fluid components, the wall can produce either attractive or repulsive gravitational field, or, in a fine tuning, even a null gravitational influence.

\section{Final Remarks}

The main result of this paper is to show that the braneworld picture can receive static domain walls, although general relativity prohibits that structures. Also, we see that static domain walls within the braneworld context can have different properties, like an attractive gravitational field. In fact, it is known that domain walls can be used to distinguish between modified gravity theories and general relativity \cite{probes}.

An interesting result of the domain wall gravitation in braneworld is that the stress-energy tensor of the wall does not have any extra influence over the gravitational dynamics if compared with Einstein's gravity. It occurs because the term $\Pi_{\mu\nu}$ on the right-hand side of (\ref{dyn}) vanishes for the domain wall stress-energy tensor. The unique extra contribution performed by the braneworld model is due to the Weyl fluid term. Meanwhile, if we consider a conformal metric for the bulk spacetime, the Weyl tensor vanishes as well as does the Weyl fluid term in equation (\ref{dyn}). In that case, we recover the result of \cite{padilla1,padilla2}, where there is no difference between Einstein's gravity and braneworld gravity for the domain wall spacetime. In the general relativity case, to probe the inconsistency of static domain walls \cite{ray}  it is considered an energy distribution over the nearly Minkowski spacetime and assumed the staticity of the spacetime into the domain wall. In this way,  it is reasonable to expect that the system no longer keeps over Minkowski spacetime, since in fact we do have gravitational source. In the braneworld context, however, the dark fluid offset the influence of the stress-energy of the domain wall in such a manner that it becomes possible that a static domain wall appears in this context.

Let us finalise saying that, in this work we have implemented the so-called brane-based formalism, in which the 3-brane is localized on a point of extra-dimension and, by means Gauss-Codazzi equations and junction conditions, it is possible to obtain the brane effective gravity. Such an approach retains undetermined the bulk contribution over the braneworld gravitation (encoded by means of the dark fluid term in Eq. (\ref{eq:c6})). That indeterminacy over the Weyl fluid form allow us to constrain some components so that it satisfies the staticity condition for the existence of static domain walls in braneworld scenarios. Therefore, this method restrict the bulk gravitation. There is another approach in which we firstly determine the bulk metric properties, the so-called bulk-based formalism. For a spherically vacuum bulk spacetime, a generalization of the Birkhoff theorem \cite{gregory2} states that the metric bulk represents a Schwarzschild-AdS spacetime. For this case, a moving brane in the bulk represents an expanding universe. Taken a bulk Schwarzschild-AdS metric, it was demonstrated \cite{RM} that the Weyl fluid components behaves as follows: the dark pressure term is null whilst the dark radiation (associated with the bulk black hole mass) is fairly constant. Therefore, in view of Eqs. (\ref{eq61}) and (\ref{eq58}), we conclude  that static domain walls are incompatible with the metrics type Schwarzschild-AdS for the bulk, since the staticity condition cannot be fulfilled in this case.

\section*{Acknowledgements}

M. C. B. Abdalla and J. M. Hoff da Silva acknowledge to Conselho Nacional de Desenvolvimento Cient\'{\i}fico e Tecnol\'ogico (CNPq) for partial financial support (482043/2011-3; 308623/2012-6). P. F. Carlesso acknowledges to CAPES-Brazil for financial suport.

\end{document}